\documentclass[11pt]{article}
\usepackage{amsmath,amsthm}

\newcommand{\R}{{\mathbf R}} \newcommand{\N}{{\mathbf N}}
\newcommand{\K}{{\mathbf K}} \newcommand{\Z}{{\mathbf Z}}
  \def\C{{\mathbf C}}
\newcommand{\Prm}{{\mathbf P}}
\newcommand{\wt}{\widetilde }

\renewcommand{\epsilon}{\varepsilon } 
\newcommand{\g}{\gamma } 
\renewcommand{\rho}{\varrho } 
\renewcommand{\l}{{\lambda }} 
\renewcommand{\phi}{\varphi }

\renewcommand{\b}{\beta }
\newcommand{\E}{{\mathbf E}}
\newcommand{\q}{{\rm q }}

\def\rs{\right>}
\def\lg{\left|}

\newtheorem{theorem}{Theorem}
\newtheorem{lemma}{Lemma}
\newtheorem{corollary}{Corollary}
\newtheorem{proposition}{Proposition}

\begin {document}
 \title{Quantum Approximation I. Embeddings of Finite Dimensional $L_p$ Spaces}

\author {Stefan Heinrich\\
Fachbereich Informatik\\
Universit\"at Kaiserslautern\\
D-67653 Kaiserslautern, Germany\\
e-mail: heinrich@informatik.uni-kl.de\\
homepage: http://www.uni-kl.de/AG-Heinrich}   
\date{}
\maketitle

\date{}
\maketitle
\begin{abstract}
%\noindent
We study  approximation of embeddings between finite dimensional
$L_p$ spaces in the quantum model of computation. 
For the quantum query complexity of this problem matching (up to logarithmic factors)
upper and lower bounds are obtained. The results show that for certain regions 
of the parameter domain  quantum computation can essentially
improve the rate of convergence of
classical deterministic or randomized approximation, while there are other regions
 where the best possible rates coincide for all three settings. These results serve as 
a crucial building block for analyzing approximation in function spaces in
a subsequent paper \cite{Hei03b}.
\end{abstract}
\section{Introduction}
In this paper we continue the investigation of numerical problems of 
analysis in the quantum model of computation. In a number of papers the integration problem
and its discretized version, the mean computation, were studied  and matching upper
and lower bounds (often up to logarithmic factors) were established for
various function classes. See the references \cite{BHMT00,G2,NW,Nov01,Hei01,HN01b,Hei01b,TW01,
KW03, HKW03}.
It turned out that for these types of problems quantum computing can reach an exponential 
speedup over deterministic classical computation and a quadratic speedup over
randomized classical computation. 

All these problems are such that the solution is a 
single number. Therefore the question arises what happens if we consider problems whose
solution is a family of numbers or, in other words, a function.  A particularly typical
situation is function approximation, where the solution is just the input function itself
(and we are asked to compute an approximation to it in a given norm). A first consideration
of an approximation problem in the quantum setting appears in \cite{NSW02}, but no matching 
upper and lower bounds were obtained.

In the present paper we provide the first results for approximation in the quantum model of
computation, with matching upper and lower bounds. 
We start with the very basic situation: the approximation of 
the embedding $J_{pq}^N$ of $L_p^N$ into
$L_q^N$, or, in other words, the approximation of $N$-sequences with bounded $L_p^N$ norm
in the norm of $L_q^N$.
These embeddings are the elementary building blocks of embeddings of function spaces
 -- in the same way as mean computation is the elementary building block of integration,
see \cite{Hei01b} and \cite{Hei03b} for more on this principle.
Our results show that for $p<q$, the quantum model of computation can bring an acceleration 
 up to a factor $N^{-1}$ of the rate of the classical (deterministic or randomized) 
setting.
On the other hand, for $p\ge q$, the optimal rate is the same for all three settings,
so in these cases there is no speedup of the rate by quantum computation.

We prove that the following version of  
 Grover's quantum search algorithm is optimal: we find all 
 coordinates of 
 $f\in L_p^N$ with absolute 
value not smaller than a suitably chosen threshold and set the other coordinates to zero.
The crucial new element in proving lower bounds is a multiplicative inequality
for the $n$-th minimal query error, which is
-- in a wide sense -- analogous to multiplicativity properties of $s$-numbers,
see \cite{Pie87}.

In a subsequent paper \cite{Hei03b} we show that, similarly to the analysis in 
\cite{Hei01b},
 sufficiently precise knowledge about the embeddings $J_{pq}^N$
leads to a full understanding of the
infinite dimensional problem of approximation of functions from Sobolev spaces.

The paper is organized as follows. In Section 2 we recall  notation 
from the quantum setting of information-based complexity theory as developed 
in \cite{Hei01}.
In Section 3 we derive some new general results which will be needed later on.
Section 4 contains the main results on approximation of embeddings
of  $L_p^N$  into $L_q^N$ spaces. Finally, in Section 5 we give some 
comments on the quantum bit model and a
summary including comparisons to the respective results in the 
classical deterministic and randomized setting.

For more details on the quantum setting of information-based complexity we refer to
\cite{Hei01}, to the survey \cite{HN01a}, and to an introduction  \cite{Hei01a}. 
For the classical settings of information-based complexity theory
see \cite{Nov88, TWW88, Hei93}.   
General background on quantum computing can be found in the surveys 
\cite{Aha98,EHI00, Sho00} and in the 
monographs \cite{Pit99, 
Gru99} and \cite{NC00}.

\section{Notation}

For nonempty sets $D$ and $K$, we denote by $\mathcal{F}(D,K)$
the set of all functions from $D$ to $K$. For a normed space $G$ we let
$\mathcal{B}(G)=\{g\in G\,|\,\|g\|_{G}\le 1\}$ denote the unit ball of $G$. Let 
$F \subseteq \mathcal{F}(D, K)$ be a nonempty subset. 
Let $\K$ stand for either 
$\R$ or $\C$, the field of real or complex numbers, let  
$G$ be a normed space over $\K$, and 
let $S:F\to G$ be a mapping. We seek to 
approximate $S(f)$ for $f\in F$ by means of quantum computations. 
Let $H_1$ be the 
two-dimensional complex Hilbert space $\C^2$, with its unit vector
basis $\{e_0,e_1\}$, let
$$
 H_m=\underbrace{H_1\otimes\dots\otimes H_1}_{m}, 
$$
equipped with the tensor
Hilbert space structure. 
Denote
$$\Z[0,N) := \{0,\dots,N-1\}$$
for $N\in\N$, where we agree to write, as usual, $\N= \{1,2,\dots \}$ 
and $\N_0=\N\cup\{0\}$.
Let $\mathcal{C}_m = \{\lg i\rs:\, i\in\Z[0,2^m)\}$ be the canonical basis of
$H_m$, where  $\lg i \rs$ stands for 
$e_{j_0}\otimes\dots\otimes e_{j_{m-1}}$, $i=\sum_{k=0}^{m-1}j_k2^{m-1-k}$ is the binary 
expansion of $i$. Let $\mathcal{U}(H_m)$ denote the set of unitary operators on $H_m$. 

A quantum query  on $F$ is given by a tuple
\begin{equation*}
Q=(m,m',m'',Z,\tau,\beta),
\end{equation*}
where $m,m',m''\in \N, m'+m''\le m, Z\subseteq \Z[0,2^{m'})$ is a nonempty 
subset, and
$$\tau:Z\to D$$
$$\beta:K\to\Z[0,2^{m''})$$
are arbitrary mappings. We let $m(Q):=m$ be the number of qubits of $Q$. 

Given a query $Q$, we define for each $f\in F$ the unitary operator 
$Q_f$ by setting for  
$\lg i\rs\lg x\rs\lg y\rs\in \mathcal{C}_m
=\mathcal{C}_{m'}\otimes\mathcal{C}_{m''}\otimes\mathcal{C}_{m-m'-m''}$:
\begin{equation*}
Q_f\lg i\rs\lg x\rs\lg y\rs=
\left\{\begin{array}{ll}
\lg i\rs\lg x\oplus\beta(f(\tau(i)))\rs\lg y\rs &\quad \mbox {if} \quad i\in Z\\
\lg i\rs\lg x\rs\lg y\rs & \quad\mbox{otherwise,} 
 \end{array}
\right. 
\end{equation*}
where $\oplus$ means addition modulo $2^{m''}$. 

A quantum algorithm  on $F$  with no measurement is a tuple
\begin{equation*}
A=(Q,(U_j)_{j=0}^n).
\end{equation*}
Here $Q$ is a quantum query on $F$, $n\in\N_0$  and
$U_{j}\in \mathcal{U}(H_m)\,(j=0,\dots,n)$, with $m=m(Q)$.
Given $f\in F$,
we define $A_f\in \mathcal{U}(H_m)$ as
\begin{equation*}
A_f = U_n Q_f U_{n-1}\dots U_1 Q_f U_0.
\end{equation*}
We denote by $n_q(A):=n$ the number of queries and by $m(A)=m=m(Q)$ the 
number of qubits of $A$. Let $(A_f(x,y))_{x,y\in \Z[0,2^m)}$ 
be the matrix of the 
transformation $A_f$ in the canonical basis $\mathcal{C}_{m}$, that is,
$A_f(x,y)=(A_f\lg y \rs,\lg x\rs)$.

A quantum algorithm from $F$ to $G$ with $k$ measurements  is a tuple
$$
A=((A_l)_{l=0}^{k-1},(b_l)_{l=0}^{k-1},\varphi),
$$ 
where $k\in\N$, $A_l\;(l=0,\dots,k-1)$ are quantum algorithms
on $F$ with no measurement, 
$$
b_0\in\Z[0,2^{m_0}), 
$$
$$
b_l:\prod_{i=0}^{l-1}\Z[0,2^{m_i}) \to \Z[0,2^{m_l})\quad 
(1\le l \le k-1),
$$
where $m_l:=m(A_l)$, and 
$$
\varphi:\prod_{l=0}^{k-1}\Z[0,2^{m_l}) \to G.
$$
The output of $A$ at input $f\in F$ is a probability measure $A(f)$ on $G$, 
defined as follows. First put
\begin{eqnarray}   
p_{A,f}(x_0,\dots, x_{k-1})&=&
|A_{0,f}(x_0,b_0)|^2 |A_{1,f}(x_1,b_1(x_0))|^2\dots\nonumber\\
&&\dots |A_{k-1,f}(x_{k-1},b_{k-1}(x_0,\dots,x_{k-2}))|^2.\nonumber
\end{eqnarray}
Then define $A(f)$ by setting for any subset $C\subseteq G$
\begin{equation*}
A(f)(C)=\sum_{\phi(x_0,\dots,x_{k-1})\in C}p_{A,f}(x_0,\dots, x_{k-1}).
\end{equation*}
Let 
$n_q(A):=\sum_{l=0}^{k-1} n_q(A_l)$
denote the number of queries used by $A$.
For more details and background see \cite{Hei01}.
Note that we often use the term `quantum algorithm' (or just `algorithm'), 
meaning a quantum algorithm with measurement(s).

If $A$ is an algorithm with one measurement, the above definition 
simplifies essentially. Such an algorithm is given by
\begin{equation}
\label{ZB1}
A=(A_0,b_0,\phi),\quad A_0=(Q,(U_j)_{j=0}^n).
\end{equation}
The quantum computation is carried out on $m:=m(Q)$ qubits. For $f\in F$ the 
algorithm starts in the state  $\lg b_0 \rs$ and produces
\begin{equation}
\label{ZA2}
\lg \psi_f \rs=U_n Q_f U_{n-1}\dots U_1 Q_f U_0\lg b_0 \rs.
\end{equation}
Let
\begin{equation}
\label{ZA3}
\lg \psi_f \rs=\sum_{i=0}^{2^m-1}a_{i,f}\lg i \rs
\end{equation}
(referring to the notation above, we have $a_{i,f}=A_{0,f}(i,b_0)$). Then $A$ outputs the 
element $\phi(i)\in G$ with probability $|a_{i,f}|^2$. It is shown in \cite{Hei01},
Lemma 1, that for each algorithm $A$ with $k$ measurements there is an algorithm 
$\wt{A}$ with one measurement such that $A(f)=\wt{A}(f)$ for all $f\in F$ and 
$\wt{A}$ uses just twice the number of queries of $A$, that is, $n_q(\wt{A})=2n_q(A)$.
Hence, as long as we are concerned with studying the minimal query error
(see below) up to the 
order, that is, up to constant factors, we can 
restrict ourselves to algorithms with one measurement.

Let $\theta\ge 0$. For a quantum algorithm $A$ we 
define the (probabilistic) error at $f\in F$ as follows. 
Let 
$\zeta$ be a random variable with distribution $A(f)$. Then 
\begin{equation*}
e(S,A,f,\theta)=\inf\left\{\varepsilon\ge0\,\,|\,\,\Prm\{\|S(f)-\zeta\|>\varepsilon\}
\le\theta
\right\}
\end{equation*}
(note that this infimum is always attained).
Hence $e(S,A,f,\theta)\le \varepsilon$ iff the probability that the algorithm $A$ 
computes $S(f)$  with 
error at most $\varepsilon$ is  at least $1-\theta$.
Observe that for algorithms with one measurement ((\ref{ZB1})-(\ref{ZA3})),
$$
   P\{\|S(f)-\zeta\|>\varepsilon\}\,=\,\sum_{i:\
   \|S(f)-\phi(i)\|>\varepsilon}|a_{i,f}|^2.
$$
Trivially, $e(S,A,f,\theta)=0$ for $\theta\ge 1$. 
We put 
$$
e(S,A,F,\theta)=\sup_{f\in F} e(S,A,f,\theta) 
$$
(we allow the value $+\infty$ for this quantity). Furthermore, we set
\begin{eqnarray*}
\lefteqn{e_n^\q(S,F,\theta)}\\
&=&\inf\{e(S,A,F,\theta)\,\,|\,\,A\,\,
\mbox{is any quantum algorithm with}\,\, n_q(A)\le n\}.
\end{eqnarray*}
It is customary to consider these quantities at a fixed error probability
level. We denote
$$
e(S,A,f)=e(S,A,f,1/4)
$$
and similarly,
$$
e(S,A,F)=e(S,A,F,1/4),\quad e_n^\q(S,F)=e_n^\q(S,F,1/4).
$$
The choice $\theta=1/4$ is arbitrary -- any fixed $\theta<1/2$ would do.
The quantity $e_n^\q(S,F)$ is  central for our study -- it is the 
$n$-th minimal query error, 
that is, the smallest error which can be reached using at most $n$ queries.
Note that it essentially suffices to study
$e_n^\q(S,F)$ instead of $e_n^\q(S,F,\theta)$, 
since with $\mathcal{O}(\nu)$ repetitions, the error probability 
can be reduced to $2^{-\nu}$ (see Lemmas \ref{lem:2e}, \ref{lem:2f}, and 
Corollary \ref{cor:4} below).

\section{Some General Results}
Let $G$ and $\wt{G}$ be normed spaces. Recall that a mapping $\Phi:G\to \wt{G}$ 
is said to be Lipschitz, if there is a constant $c\ge 0$ such that
$$
\|\Phi(x)-\Phi(y)\|_{\wt{G}}\le c\,\|x-y\|_G \quad\mbox{for all}\quad x,y\in G.
$$ 
The Lipschitz constant $\|\Phi\|_{\rm Lip}$ is the smallest constant $c$ such that
the relation above holds.

Given a quantum algorithm $A$ from $F$ to $G$ and a mapping $\Phi$ from
$G$ to $\wt{G}$, the algorithm $\Phi\circ A$ is defined as the composition, meaning that
$\phi$ in the definition of $A$ is replaced by $\Phi\circ\phi$ (this is a special case
of the definition of  the composition 
given in \cite{Hei01}, p.\ 13). The following direct consequence of the
definitions will be needed later.
\begin{lemma}
\label{lem:lip}
Let $S$ be a mapping  and $A$ a quantum algorithm, both from $F$ to $G$. Let $\Phi$
be a Lipschitz mapping from
$G$ to $\wt{G}$. Then for each $f\in F$ and $\theta\ge 0$,
$$
e(\Phi\circ S,\Phi\circ A,f,\theta)\le \|\Phi\|_{\rm Lip} e(S,A,f,\theta).
$$
Consequently, for each $n\in\N$,
$$
e_n^\q(\Phi\circ S,F,\theta)\le \|\Phi\|_{\rm Lip} e_n^\q(S,F,\theta).
$$
\end{lemma}
The next result was shown in \cite{Hei01b}, Lemma 2, for $G=\R$, but 
the proof of the general case is identical to that one.

\begin{lemma}
\label{lem:2}
Let $D,K$ and $F\subseteq\mathcal{F}(D,K)$ be nonempty sets, G a normed space,
let $k\in\N_0$
and let $S_l:F\to G$ $(l=0,\dots,k)$ be mappings. Define $S:F\to G$ by
$S(f)=\sum_{l=0}^k S_l(f)\quad(f\in F)$. Let $\theta_0,\dots,\theta_k\ge 0$,
$n_0,\dots,n_k\in\N_0$ and put $n=\sum_{l=0}^k n_l$.   Then
$$
e_n^\q(S,F,\sum_{l=0}^k\theta_l)\le\sum_{l=0}^k e_{n_l}^\q(S_l,F,\theta_l).
$$
\end{lemma}

The following results are generalizations of the usual procedure of 
"boosting the success probability", which decreases the failure probability by 
repeating the algorithm a number of times and computing the median of the outputs
(see, e.g., \cite{Hei01}, Lemma 3). This works for algorithms whose outputs are
real numbers. 
Since there is no natural linear order on a normed space, in general, 
the latter step has to be changed suitably when dealing with outputs in a 
normed space $G$.

For this purpose, let  $\nu\in \N$. Let 
$\mu:\R^\nu\to \R$ be the mapping given by the median, that is, 
$\mu(a_0,\dots,a_{\nu-1})$ is the value 
of the $\lceil (\nu+1)/2\rceil$-th element of the
non-decreasing rearrangement of $(a_i)$. First we deal with the case that $G$ is
a space of the form $G=l_\infty(\mathcal{T})$, where $\mathcal{T}$ 
is a nonempty set and
$l_\infty(\mathcal{T})$ denotes the space of all bounded real-valued functions 
on $\mathcal{T}$, equipped with the supremum norm 
$\|g\|_{l_\infty(\mathcal{T})}=\sup_{t\in \mathcal{T}}|g(t)|$.
Define $\bar{\mu}:l_\infty(\mathcal{T})^\nu\to l_\infty(\mathcal{T})$ as follows: 
$$
\bar{\mu}(g_0,\dots,g_{\nu-1})=(\mu(g_0(t),\dots,g_{\nu-1}(t)))_{t\in \mathcal{T}},
$$
that is, we apply the median componentwise.
 For any algorithm $A$ from $F$ to $l_\infty(\mathcal{T})$ denote
by $\bar{\mu}(A^\nu):= \bar{\mu}(A,\dots,A)$ the composed algorithm 
(see again p.\ 13 of 
\cite{Hei01})) of repeating $\nu$ times the algorithm $A$  and applying $\bar{\mu}$ to 
the outputs.
\begin{lemma}
\label{lem:2e}
Let $A$ be any quantum algorithm and $S$ be any mapping, both from $F$ to 
$l_\infty(\mathcal{T})$, and let $\nu\in\N$. 
Then for each $f\in F$,
$$
e(S,\bar{\mu}(A^\nu),f,e^{-\nu/8})\le e(S,A,f).
$$
\end{lemma}
\begin{proof}
Fix $f\in F$.
Let $\zeta_0,\dots,\zeta_{\nu-1}$ be independent random variables with distribution $A(f)$.
Let $\chi_i$ be the indicator function of the set
$$
\{\|S(f)-\zeta_i\|_{l_\infty(\mathcal{T})}>e(S,A,f)\}.
$$
Then $\Prm\{\chi_i=1\}\le 1/4$. Hoeffding's inequality, see, e.g., \cite{Pol}, p.\ 191,
yields
$$
\Prm\left\{\sum_{i=0}^{\nu-1}\chi_i\ge \nu/2\right\}\le
\Prm\left\{\sum_{i=0}^{\nu-1}(\chi_i-\E\chi_i)\ge \nu/4\right\}\le e^{-\nu/8}.
$$
Hence, with probability at least $1-e^{-\nu/8}$,   
\begin{equation}
\label{AE1}
|\{i\,\,|\,\,\|S(f)-\zeta_i\|_{l_\infty(\mathcal{T})}\le e(S,A,f)\}|> \nu/2.
\end{equation}
It follows from (\ref{AE1}) that for all $t\in \mathcal{T}$
$$
|\{i\,\,|\,\,|S(f)(t)-\zeta_i(t)|\le e(S,A,f)\}|> \nu/2.
$$
Consequently,
$$
|S(f)(t)-\mu(\zeta_0(t),\dots,\zeta_{\nu-1}(t))|\le e(S,A,f),
$$
which means that
\begin{equation}
\label{AE2}
\|S(f)-\bar{\mu}(\zeta_0,\dots,\zeta_{\nu-1})\|_{l_\infty(\mathcal{T})}\le e(S,A,f).
\end{equation}
Since (\ref{AE1}) holds with probability at least $1-e^{-\nu/8}$, so does
(\ref{AE2}).
\end{proof}
Now let $G$ be a general normed space. For the following construction
we consider $G$ as a space over $\R$ (each normed space over $\C$ can also be considered as
a normed space over $\R$).  We define for each $\delta>0$ a suitable 
mapping $\psi_\delta:G^\nu\to G$ as follows. Let $G^*$ denote the dual of $G$, that is,
the space of all bounded linear functionals on $G$. Let 
$\mathcal{T}\subseteq \mathcal{B}(G^*)$
be a norming set, i.e., for all $g\in G$
\begin{equation}
\label{AE3}
\|g\|=\sup_{t\in \mathcal{T}}|t(g)|
\end{equation}
(such a $\mathcal{T}$ always exists, one can take, for example,
$\mathcal{T}= \mathcal{B}(G^*)$ itself). Then $G$ can be identified with a subspace 
of $l_\infty(\mathcal{T})$ via the embedding map $J:G\to l_\infty(\mathcal{T})$ 
defined by
$$
Jg=(t(g))_{t\in \mathcal{T}}
$$
in such a way that the norm is preserved: $\|Jg\|_{l_\infty(\mathcal{T})}=\|g\|_G$.
For $\delta>0$, let finally  $\pi_\delta:l_\infty(\mathcal{T})\to G$ be any 
$\delta$-approximate
metric projection, by which we mean a mapping satisfying
$$
\|x-\pi_\delta(x)\|_{l_\infty(\mathcal{T})}
\le (1+\delta)\inf_{g\in G}\|x-g\|_{l_\infty(\mathcal{T})}
$$
for all $x\in l_\infty(\mathcal{T})$. We define $\psi_\delta:G^\nu\to G$ by setting
$$\psi_\delta=\pi_\delta\circ\bar{\mu}\circ J^\nu.$$
\begin{lemma}
\label{lem:2f}
Let $A$ be any quantum algorithm and $S$ be any mapping, both from $F$ to a normed space
$G$, let $\nu\in\N$ and $\delta>0$. 
Then for each $f\in F$,
$$
e(S,\psi_\delta(A^\nu),f,e^{-\nu/8})\le (2+\delta)e(S,A,f).
$$
\end{lemma}
\begin{proof}
It follows from (\ref{AE3}) that 
$$
e(JS,JA,f)=e(S,A,f).
$$
By Lemma \ref{lem:2e}, 
$$
e(JS,\bar{\mu}((JA)^\nu),f,e^{-\nu/8})\le e(JS,JA,f)=e(S,A,f).
$$
Let $\zeta$ be a random variable with values in $l_\infty(\mathcal{T})$ with 
distribution
$\bar{\mu}((JA)^\nu)(f)$. Then with probability at least $1-e^{-\nu/8}$,
$$
\|JS(f)-\zeta\|\le e(S,A,f).
$$
Hence 
\begin{eqnarray*}
\|S(f)-\pi_\delta(\zeta)\|&\le& \|JS(f)-\zeta\|+\|\zeta-\pi_\delta(\zeta)\|\\
&\le& e(S,A,f)+(1+\delta)\|\zeta-S(f)\|\\
&\le& (2+\delta)e(S,A,f).
\end{eqnarray*}
But $\pi_\delta(\zeta)$ is a random variable with distribution 
$$
\pi_\delta\circ\bar{\mu}((JA)^\nu)(f)=\psi_\delta(A^\nu)(f),
$$
and it follows that 
$$
e(S,\psi_\delta(A^\nu),f,e^{-\nu/8})\le (2+\delta)e(S,A,f).
$$ 
 
\end{proof}
\begin{corollary}
\label{cor:4}
Let $S$ be any mapping from $F\subseteq\mathcal{F}(D,K)$ to a normed space $G$. 
Then for each $n,\nu\in\N$,
$$
e_{\nu n}^\q(S,F,e^{-\nu/8})\le 2e_n^\q(S,F).
$$
If $G$ is a space of the form $l_\infty(\mathcal{T})$ for some set $\mathcal{T}$, 
then the constant $2$ above can be replaced by $1$.
\end{corollary}
The definition of $\psi_\delta$ is not constructive (and neither is that of $\bar{\mu}$,
if $\mathcal{T}$ is infinite).
 Since we are dealing with the quantum query complexity, 
the cost of
(classically) computing the $\phi$ part of a quantum algorithm 
(this is the place where $\bar{\mu}$ and $\psi_\delta$ enter)
 are generally neglected. However, if one looks for a more efficient 
procedure, here is one which leads to the constant 3 instead of $2+\delta$ of Lemma
\ref{lem:2f}.
Define $\rho:G^\nu\to G$ as follows: $\rho(g_0,\dots,g_{\nu-1})$
is the element $g_{i_0}$, where 
$$
i_0= \mbox{arg}\min_i \mu(\|g_0-g_i\|,\dots,\|g_{\nu-1}-g_i\|)
$$
(if there is more than one index $i$ at which the minimum is attained, we choose
the smallest index, just for definiteness).
It can be shown along the lines of the proof of Lemma \ref{lem:2e} that 
for all $f\in F$,
 $$
e(S,\rho(A^\nu),f,e^{-\nu/8})\le 3e(S,A,f).
$$
Note that the cost is $\mathcal{O}(\nu^2)$ (which is usually a logarithmic term) times the cost of
computing the norm  $\|g_i-g_j\|$ (which depends on the structure and dimension of $G$,
 and on the -- possible -- sparsity
of the $g_i$, see also the discussion in Section 3 of \cite{Hei03b}).

\begin{corollary}
\label{cor:3}
Let $D,K$,  $F\subseteq\mathcal{F}(D,K)$, G, $k\in\N_0$
and $S,S_l:F\to G$ $(l=0,\dots,k)$ be as in Lemma \ref{lem:2}. 
 Assume $\nu_0,\dots,\nu_k\in \N$ satisfy
$$
\sum_{l=0}^k e^{-\nu_l/8}\le \frac{1}{4}.
$$ 
Let $n_0,\dots,n_k\in\N_0$ and put
$n=\sum_{l=0}^k \nu_l n_l$. Then
$$
e_n^\q(S,F)\le2\sum_{l=0}^k e_{n_l}^\q(S_l,F).
$$
If $G=l_\infty(\mathcal{T})$, then the relation holds with
constant $1$.
\end{corollary}
This is an obvious consequence of Lemma \ref{lem:2}  and Corollary \ref{cor:4}. 
In the sequel we need the following mappings. Let $m^*\in \N$ and define
$\b:\R\to\Z[0,2^{m^*})$ for $z\in\R$ by
\begin{equation}\label{N1}
\b(z)=
\left\{\begin{array}{lll}
   0& \mbox{if} \quad z <-2^{m^*/2-1} \\
   \lfloor 2^{m^*/2}(z+2^{m^*/2-1})\rfloor       & \mbox{if} \quad  
   -2^{m^*/2-1}\le z <2^{m^*/2-1}\\
   2^{m^*}-1& \mbox{if} \quad z\ge 2^{m^*/2-1}. 
   \end{array}
   \right.
\end{equation}
Furthermore, let $\gamma:\Z[0,2^{m^*})\to\R$ be defined for $y\in\Z[0,2^{m^*})$ as 
\begin{equation}
\label{N2}
\g(y)=2^{-m^*/2}y-2^{m^*/2-1}.
\end{equation}
It follows that for $-2^{m^*/2-1}\le z\le 2^{m^*/2-1}$,
\begin{equation}
\label{E4}
\g(\b(z))\le z\le \g(\b(z))+2^{-m^*/2}.
\end{equation}
\begin{proposition}
\label{pro:4} 
Let $D$ be a nonempty set and let 
$\emptyset\ne F\subseteq X \subseteq\ \mathcal{F}(D,\R)$, where $X$ is a linear 
subspace equipped
with a norm $\|\:\|_X$, such that\\
(i) $\sup_{f\in F}|f(t)|<\infty$ for each $t\in D$,  and\\ 
(ii) $X$ separates the points of $D$ in the following sense: Given $t_0\in D$ 
and a finite subset $D_0\subseteq D\backslash\{t_0\}$, there is an $g\in X$ with 
$g(t_0)\ne 0$ and $g(t)=0$ for all $t\in D_0$.

Let $J:F\to X$ be the embedding map, let $G$ be a normed space and  
$S:X\to G$ a bounded linear operator. Then
for all $\tilde{n}, n\in \N$, $0\le\theta_1,\theta_2\le1$,
$$
e^\q_{\tilde{n}+2n}(SJ,F,\theta_1+\theta_2-\theta_1\theta_2)
\le 
e^\q_{\tilde{n}}(J,F,\theta_1)\,e^\q_{n}(S,\mathcal{B}(X),\theta_2).
$$
\end{proposition}
\begin{proof}
Let $\delta>0$, let $\wt{A}$ be a quantum algorithm from $F$ 
to $X$ with $q(\wt{A})\le \tilde{n}$ and
\begin{equation}
\label{AB7a}
e(J,\wt{A},F,\theta_1)
\le e^\q_{\wt{n}}(J,F,\theta_1)+\delta:=\sigma_1.
\end{equation}
Put
\begin{equation}
\label{AC1}
\sigma=\sigma_1+\delta.
\end{equation}
Let $A$ be a quantum algorithm from $\mathcal{B}(X)$ to $G$ with $q(A)\le n$
and 
\begin{equation}
\label{AB7b}
e(S,A,\mathcal{B}(X),\theta_2)
\le e_n^\q(S,\mathcal{B}(X),\theta_2)+\delta:=\sigma_2.
\end{equation}
Let 
$$
\wt{A}=((\wt{A}_l)_{l=0}^{\tilde{k}-1},(\wt{b}_l)_{l=0}^{\tilde{k}-1},\wt{\varphi}),
$$
with
$$
\wt{A}_l=(\wt{Q}_l,(\wt{U}_{l j})_{j=0}^{\wt{n}_l}),
$$
and
$$
\wt{Q}_l=(\wt{m}_l,\wt{m}'_l,\wt{m}''_l,\wt{Z}_l,
\wt{\tau}_l,\wt{\beta}_l).
$$
for $l=0,\dots,\tilde{k}-1$.
Furthermore, let
$$
A=((A_l)_{l=0}^{k-1},(b_l)_{l=0}^{k-1},\varphi),
$$
and for $l=0,\dots,k-1$,
$$
A_l=(Q_l,(U_{l j})_{j=0}^{n_l}),
$$
and
$$
Q_l=(m_l,m'_l,m''_l,Z_l,\tau_l,\beta_l).
$$
We need some auxiliary functions and relations. Let 
$$
D_A=\{\tau_l(i)\,|\,l=0,\dots,k-1,\, i\in Z_l\}
$$
(the set of all points at which the quantum algorithm $A$ queries the function).
By assumption (ii), for each $t\in D_A$ there is a $g_t\in X$ such that 
$g(t)=1$ and $g(s)=0$ for all $s\in D_A\backslash\{t\}$.
Let $M_1=\max_{t\in D_A}\|g_t\|_X$. By asumption (i), 
$$
M_2:=\max_{f\in F,\, t\in D_A}|f(t)|<\infty.
$$
Now choose the $m^*$ in the definition of the mappings $\b$, $\g$ in (\ref{N1}) and
(\ref{N2}) in such a way that for $a\in \R$ with $|a|\le M_2$,
$$
|a-\g\circ \beta(a)|\le M_1^{-1}|D_A|^{-1}\delta.
$$
Define for $f\in F$ and 
$x=(x_0,\dots,x_{\tilde{k}-1})\in \prod_{l=0}^{\tilde{k}-1}\Z[0,2^{\wt{m}_l})$,
$$
h_{f,x}=\sigma^{-1}\left(f-\wt{\phi}(x)+\sum_{t\in D_A}(\g\circ\b\circ f(t)-f(t))g_t\right)
$$
(recall that $\wt{\phi}(x)\in X$). Then $h_{f,x}\in X$, 
\begin{equation}
\label{AC5}
h_{f,x}(s)=\sigma^{-1}(\g\circ\b\circ f(s)-\wt{\phi}(x)(s))\quad (s\in D_A),
\end{equation}
and
\begin{equation}
\label{AC3}
\|f-\wt{\phi}(x)-\sigma h_{f,x}\|_X\le M_1|D_A|M_1^{-1}|D_A|^{-1}\delta=\delta .
\end{equation}
Moreover,
\begin{eqnarray}
\|h_{f,x}\|_X&=&\sigma^{-1}\|(f-\wt{\phi}(x))-(f-\wt{\phi}(x)-\sigma h_{f,x})\|_X\nonumber\\
&\le& \sigma^{-1}(\|(f-\wt{\phi}(x)\|_X+\delta).  \label{AC7}
\end{eqnarray}
We build an algorithm as follows. It has $\tilde{k}+k$ cycles. The first 
$\tilde{k}$ cycles are exactly those of $\wt{A}$. After the $\tilde{k}$ measurements we
have the result, say
$$
x=( x_0,x_1,\dots, x_{\tilde{k}-1}),
$$
(from which $\wt{\varphi}(x_0,\dots,x_{\tilde{k}-1})$ would be computed -- 
but we don't do that yet). Next the $k$ cycles of $A$ follow, with
certain modifications. 
In each cycle we add  $\wt{m}=\sum_{l=0}^{\tilde{k}-1}\wt{m}_l$
qubits which are initialized in the state 
$$
\lg x \rs=\lg x_0\rs\lg x_1\rs\dots\lg x_{\tilde{k}-1}\rs
$$
and remain there all the way. We add $m^*$
further auxiliary qubits, initially set to zero (and being zero again at the 
end of each cycle).  
We also want to modify the queries $Q_l$ of $A$. For $0\le l< k$   introduce the 
following new query:
$$
\bar{Q}_l=(m_l+\wt{m}+m^*,m'_l,m^*,Z_l,\tau_l,\beta),
$$
where $\b$ was defined in (\ref{N1}). Define a unitary operator 
$V_l$ on
$$
H_{m'_{l}}\otimes H_{m''_{l}}\otimes H_{m_{l}-m'_{l}-m''_{l}}
\otimes H_{\wt{m}} \otimes H_{m^*}
$$
by setting for
$$
 \lg i \rs\lg z \rs\lg u\rs\lg x\rs\lg v\rs\in 
\mathcal{C}_{m'_{l}}\otimes \mathcal{C}_{m''_{l}}\otimes 
\mathcal{C}_{m_{l}-m'_{l}-m''_{l}}
\otimes \mathcal{C}_{\wt{m}} \otimes \mathcal{C}_{m^*}
$$
$$
V_l\lg i \rs\lg z \rs\lg u\rs\lg x\rs\lg v\rs
= \lg i \rs\lg z\oplus\b_l\Big(\sigma^{-1}\big(\g(v)-\wt{\phi}(x)
(\tau_l(i))\big)\Big)\rs\lg u\rs\lg x\rs\lg v\rs
$$
if $i\in Z_l$, and 
$$
V_l\lg i \rs\lg z \rs\lg u\rs\lg x\rs\lg v\rs
=\lg i \rs\lg z \rs\lg u\rs\lg x\rs\lg v\rs
$$
otherwise (recall that $\wt{\phi}(x)\in X\subseteq \mathcal{F}(D,\R)$ and 
$\tau_l(i)\in D$, so the respective expression above is well-defined). 
We also need 
$$
W_l\lg i \rs\lg z \rs\lg u\rs\lg x\rs\lg v\rs=
\lg i \rs\lg z \rs\lg u\rs\lg x\rs\lg \ominus v\rs
$$
with $\ominus v=(2^{m^*}-v)\mod 2^{m^*}$.
Now we consider the following composition
\begin{equation}
\label{AC4}
P_l\bar{Q}_{l,f}P_l W_l V_l P_l\bar{Q}_{l,f} P_l,
\end{equation}
where $P_l$ exchanges the $\lg z \rs$ with the $\lg  v\rs$ component. 
Let us look how the combination (\ref{AC4}) acts, when the last 
$m^*$ qubits are in the state $\lg 0\rs$. Assume $i\in Z_l$. Then
$$
\lg i \rs\lg z \rs\lg u\rs\lg x\rs\lg 0\rs
$$ 
is mapped by $P_l\bar{Q}_{l,f}P_l$ to 
$$
\lg i \rs\lg z \rs\lg u\rs\lg x\rs\lg \b\circ f\circ\tau_l(i)\rs.
$$
Using (\ref{AC5}), we see that $V_l$ produces 
\begin{eqnarray*} 
&&\lg i \rs\lg z\oplus \b_l\Big(\sigma^{-1}\big(\g\circ \b \circ f \circ\tau_l(i)
-\wt{\phi}(x)(\tau_l(i))\big)\Big)\rs\lg u\rs\lg x\rs\lg \b\circ f\circ\tau_l(i)\rs\nonumber\\
&=&\lg i \rs\lg z\oplus \b_l\circ h_{f,x}\circ\tau_l(i)
\rs\lg u\rs\lg x\rs\lg \b\circ f\circ\tau_l(i)\rs.
\end{eqnarray*}
Finally, the application of $P_l\bar{Q}_{l,f}P_l W_l$ leads to
\begin{eqnarray}
\lg i \rs\lg z\oplus \b_l\circ h_{f,x}\circ\tau_l(i)
\rs\lg u\rs\lg x\rs\lg 0\rs
&=&(Q_{l,h_{f,x}}\lg i \rs\lg z\rs\lg u\rs)\lg x\rs\lg 0\rs.\label{AC6}
\end{eqnarray}
It can be checked analogously that if $i\not\in Z_l$, we also end in the 
state given by the right-hand side of (\ref{AC6}). That means, the combination (\ref{AC4}) 
acts as if we apply the original query $Q_l$, but with $f$ replaced by $h_{f,x}$. Now we replace
each occurrence of $Q_{l,f}$ by this string (\ref{AC4}), while the $U_{l j}$ are
replaced by $\bar{U}_{l j}$, which are the $U_{l j}$,  
extended to $H_m\otimes H_{\wt{m}}\otimes H_{m^*}$ by tensoring with the
identity on $H_{\wt{m}}\otimes H_{m^*}$. The respective $\bar{b}_l$ are defined in such
a way that
$$
\bar{b}_l(x_0,\dots,x_{\tilde{k}-1}, (y_0,x,0),\dots,(y_{l-1},x,0))
=(b_l(y_0,\dots,y_{l-1}),x,0)
$$
for all $x_0,\dots,x_{\tilde{k}-1},y_0,\dots,y_{l-1}$, $l=0,\dots, k-1$.

After the completion of the $\tilde{k}+k$ cycles, let the measurement results be 
$$
x_0,\dots,x_{\tilde{k}-1}, (y_0,x,0),\dots,(y_{k-1},x,0),
$$
where, as before, $x=(x_0,\dots,x_{\tilde{k}-1})$. 
Then we apply the mapping $\bar{\phi}$ defined by
$$
\bar{\phi}(x, (y_0,x,0),\dots,(y_{k-1},x,0)):=S\wt{\phi}(x)
+\sigma\phi(y_0,\dots,y_{k-1}).
$$
Denote the resulting quantum algorithm from $F$ to $G$ by $B$.
Clearly,
\begin{equation}
\label{AC11}
q(B)=\tilde{n}+2n.
\end{equation}
By (\ref{AC6}), the modified $A$-part,
applied to $f\in F$, acts like algorithm A, applied to $h_{f,x}$. More 
precisely, in algorithm $B$, applied to $f$, given $x$ as the outcome of the measurements 
of the first part of $B$, 
the probability of measuring 
$$
(y_0,x,0),\dots,(y_{k-1},x,0)
$$
in the second part of $B$ is the same as that of measuring
$$
y_0,\dots,y_{k-1}
$$
in algorithm $A$, applied to $h_{f,x}$. For a fixed $f\in F$, we have, by 
(\ref{AB7a}), with probability at least $1-\theta_1$,
$$
\|Jf-\wt{\phi}(x)\|_X\le \sigma_1,
$$
thus, by (\ref{AC7}) (recalling also $Jf=f$),
$$
\|h_{f,x}\|_X\le\sigma^{-1}(\sigma_1+\delta)=1.
$$
Thus, for fixed $f\in F$, with probability at least $1-\theta_1$,
\begin{equation}
\label{AC9}
h_{f,x}\in \mathcal{B}(X).
\end{equation}
But for each $x$ satisfying (\ref{AC9}), we have by 
(\ref{AB7b}), with probability at least $1-\theta_2$,
\begin{equation}
\label{AC10}
\|Sh_{f,x}-\phi(y_0,\dots,y_{k-1})\|_G\le \sigma_2.
\end{equation}
Summarizing, we see that (\ref{AC9}) and (\ref{AC10}) together hold with probability
at least $(1-\theta_1)(1-\theta_2)$.
We have 
\begin{eqnarray*}
&&\|SJf-\bar{\phi}(x,(y_0,x,0),\dots,(y_{k-1},x,0))\|_G\\
&=& \|SJf-S\wt{\phi}(x)-\sigma\phi(y_0,\dots,y_{k-1})\|_G\\
&=& \|SJf-S\wt{\phi}(x)-\sigma Sh_{f,x}+\sigma Sh_{f,x}-\sigma\phi(y_0,\dots,y_{k-1})\|_G\\
&\le&\|S(f-\wt{\phi}(x)-\sigma h_{f,x})\|_G
+\sigma \|Sh_{f,x}-\phi(y_0,\dots,y_{k-1})\|_G\\
&\le&\|S\|\delta+\sigma \sigma_2,
\end{eqnarray*}
by (\ref{AC3}) and  (\ref{AC10}), with probability at least $(1-\theta_1)(1-\theta_2)$.
Thus, using (\ref{AC1}), 
$$
e(SJ,B,F, \theta_1+\theta_2-\theta_1\theta_2) 
\le \|S\|\delta+(\sigma_1+\delta)\sigma_2,
$$
hence, by (\ref{AC11}), (\ref{AB7a}), and (\ref{AB7b}),
\begin{eqnarray*}
\lefteqn{
e^\q_{\tilde{n}+2n}(SJ,F, \theta_1+\theta_2-\theta_1\theta_2)}\\
&\le& \|S\|\delta+(e^\q_{\tilde{n}}(J,F, \theta_1)+2\delta)
(e_n^\q(S,\mathcal{B}(X), \theta_2)+\delta).
\end{eqnarray*}
Since $\delta>0$ was arbitrary, the result follows.
\end{proof}
\begin{corollary}\label{cor:6}
Let $\nu_1,\nu_2\in \N$ with
$$
e^{-\nu_1/8}+e^{-\nu_2/8}-e^{-(\nu_1+\nu_2)/8}\le 1/4.
$$
Then under the same assumptions as in Proposition \ref{pro:4},
$$
e^\q_{\nu_1\tilde{n}+2\nu_2 n}(SJ,F)
\le 
4\,e^\q_{\tilde{n}}(J,F)\,e^\q_{n}(S,\mathcal{B}(X)).
$$
\end{corollary}
\begin{proof}
By Proposition \ref{pro:4} and Corollary \ref{cor:4},
\begin{eqnarray*}
%\lefteqn{
e^\q_{\nu_1\tilde{n}+2\nu_2 n}(SJ,F)
%}\\
&\le&
e_{\nu_1 \tilde{n}}^\q(J,F,e^{-\nu_1/8})
\,e_{\nu_2 n}^\q(S,\mathcal{B}(X),e^{-\nu_2/8})\\
&\le& 4\,e_{ \tilde{n}}^\q(J,F)
\,e_{n}^\q(S,\mathcal{B}(X)).
\end{eqnarray*}
\end{proof}

\section{Approximation of Finite Dimensional Embeddings}
 For $N\in\N$ and  $1\le p\le\infty$, 
let $L_p^N$ denote the space of all functions
$f:\Z[0,N)\to \R$, equipped with the norm 
$$
\|f\|_{L_p^N}=\left(\frac{1}{N}\sum_{i=0}^{N-1}|f(i)|^p \right)^{1/p}
$$
if $p<\infty$
and
$$
\|f\|_{L_\infty^N}=\max_{0\le i\le N-1} |f(i)|.
$$
Define $J^N_{pq}:L_p^N\to L_q^N$ to be the identity operator 
$J^N_{pq}f=f\;\,(f\in L_p^N) $.
Furthermore, for a real $M\ge 0$ define the operator 
$C_{pq}^{N,M}: L_p^N\to L_q^N$ for $f=(f(i))_{i=0}^{N-1}$ as
 \[
(C_{pq}^{N,M}f)(i)=\left\{\begin{array}{lll}
  f(i) & \mbox{if} \quad |f(i)|\ge M   \\
  0 & \mbox{otherwise.}    \\
    \end{array}
\right. 
\] 
\begin{lemma}
\label{lem:7} Let $1\le p,q\le\infty$. There is a constant $c>0$ sucht that
for all $n,N\in\N$, and $M\in \R$ with $M\ge 0$,
$$
e_n^\q(C^{N,M}_{pq},\mathcal{B}(L_p^N))=0
$$
whenever
$$
M\ge c(N/n)^{2/p}\max(\log(n/\sqrt{N}),1)^{2/p}.
$$
\end{lemma}
\noindent {\bf Remark.} 
Throughout the paper $\log$ means $\log_2$. Furthermore, we often use the 
same symbol $c,c_1,\dots$ for possibly different
positive constants (also when they appear in a sequence of relations).
These constants are either absolute or may depend only on $p$ and $q$ -- 
in all statements of lemmas, propositions, etc. this is precisely described anyway 
by the order of the quantifiers.
\begin{proof}
This is an immediate consequence of the proof of Proposition 1 and 
Corollary 3 of \cite{HN01b}. Namely, it contains an algorithm with 
$n$ queries that produces, with probability $\ge 3/4$, all indices $i$
with $|f(i)|\ge M$ and for each such $i$ an (arbitrarily precise) approximation
$y_i$ to $f(i)$, where $M$ is any number satisfying 
$$
M\ge c(N/n)^{2/p}\max(\log(n/\sqrt{N}),1)^{2/p}.
$$\end{proof}
\begin{lemma}
\label{lem:8} Let $1\le p\le q\le\infty$. For all $N\in\N$, 
and $M\in \R$ with $M\ge 0$,
$$
\sup_{f\in \mathcal{B}(L_p^N)}\|f-C^{N,M}_{pq}f\|_{L_q^N}\le M^{1-p/q}.
$$
\end{lemma}
\begin{proof}
We have for $f\in \mathcal{B}(L_p^N)$
\begin{eqnarray*}
\lefteqn{\frac{1}{N}\sum_{i=0}^{N-1}|f(i)-(C^{N,M}_{pq}f)(i)|^q }&&\\
&\le&\frac{1}{N}\max_j |f(j)-(C^{N,M}_{pq}f)(j)|^{q-p}
\sum_{i=0}^{N-1}|f(i)-(C^{N,M}_{pq}f)(i)|^p
\\
&\le&\frac{M^{q-p}}{N}\sum_{i=0}^{N-1}|f(i)|^p  \le M^{q-p}.
\end{eqnarray*}
\end{proof}
Next we give an upper bound. 
\begin{proposition}
\label{pro:2} 
Let $1\le p, q \le \infty$. In the case $p<q$ there is a constant $c>0$ such that 
for all $n,N\in\N$ 
$$
e_n^\q(J_{pq}^{N},\mathcal{B}(L_p^N))\le 
c\,\min\left(\left(\frac{N}{n}\log \left(n/\sqrt{N}+2\right)\right)^{2/p-2/q},N^{1/p-1/q}
\right).
$$
In the case $p\ge q$, we have
$$
e_n^\q(J_{pq}^{N},\mathcal{B}(L_p^N))\le 1.
$$
\end{proposition}
\begin{proof}
For $p<q$ the estimate involving the first term of the minimum 
follows from the previous two 
lemmas, since by Lemma 6 (i) of \cite{Hei01},
$$
e_n^\q(J_{pq}^{N},\mathcal{B}(L_p^N))\le 
e_n^\q(C_{pq}^{N,M},\mathcal{B}(L_p^N))+
\sup_{f\in \mathcal{B}(L_p^N)}\|f-C^{N,M}_{pq}f\|_{L_q^N}.
$$
The estimate involving the second term is a trivial consequence of 
$\|J_{pq}^{N}\|=N^{1/p-1/q}$. 
The case $p\ge q$ follows from $\|J_{pq}^{N}\|=1$.
\end{proof}
Before we derive lower bounds we recall some tools from \cite{Hei01}.
Let $D$ and $K$ be nonempty sets, let
$L\in \N$, and let to each $u=(u_0,\dots,u_{L-1})\in\{0,1\}^L$ an 
$f_u\in \mathcal{F}(D,K)$ be assigned such that the following 
is satisfied:
\\ \\
{\bf Condition (I):} For each $t\in D$ there is an $l$, $0\le l\le L-1$, 
such that $f_u(t)$ depends only on $u_l$, in other words, for $u,u'\in\{0,1\}^L$, 
$u_l=u'_l$ implies $f_u(t)=f_{u'}(t)$.
\\ \\ 
For $u\in\{0,1\}^L$ let $|u|$ denote the number of 1's in $u$. 
Define the function $\rho (L,l,l')$
for $L\in\N$, $0\le l\ne l'\le L$ by
\begin{equation}
\label{ACC1}
\rho (L,l,l')=\sqrt{\frac{L}{|l-l'|}}+
\frac{\min_{j=l,l'}\sqrt{j(L-j)}}{|l-l'|}.
\end{equation}
The following was proved in \cite{Hei01}, using the polynomial method
\cite{BBC:98}  and based on a result from
\cite{NW}:
\begin{lemma}
\label{lem:9} There is a constant $c_0>0$ such that the following holds:
Let $D,K$ be nonempty sets, let 
$F\subseteq\mathcal{F}(D,K)$ be a set of functions, $G$ a normed space,
$S:F\to G$ a mapping, and $L\in\N$. Suppose 
$(f_u)_{u\in\{0,1\}^L}\subseteq\mathcal{F}(D,K)$ is a system of functions satisfying
condition (I). Let finally $0\le l\ne l'\le L$ and assume that 
\begin{equation}
\label{AC2}
f_u\in F \quad {\rm whenever} \quad
|u|\in\{l,l'\}.
\end{equation}
 Then
\begin{equation}
\label{C3}
e_n^\q(S,F)\ge \frac{1}{2}\min\big\{ \|S(f_u)-S(f_{u'})\|\,\big |\, |u|=l,\, 
|u'|=l'\big\}
\end{equation}
for all $n$ with
\begin{equation}
\label{AC8}
n\le c_0\rho (L,l,l').
\end{equation}
\end{lemma}
For the case $q=\infty$ we give the following lower bound.
\begin{proposition}
\label{pro:3} 
Let $1\le p\le  \infty$. There are constants $c_1,c_2>0$ such that 
for all $n,N\in\N$ with $n\le c_1 N$ 
$$
e_n^\q(J_{p,\infty}^{N},\mathcal{B}(L_p^N))\ge 
c_2\,\min\left(\left(\frac{N}{n}\right)^{2/p},N^{1/p}
\right).
$$
\end{proposition}

\begin{proof} 
First we assume 
\begin{equation}
\label{AD1}
n\le c_0\sqrt{N},
\end{equation}
where $c_0$ is the constant from Lemma \ref{lem:9}.
Put 
$$
L=N,\quad l=0,\quad l'=1.
$$  
Then
\begin{equation}
\label{D3}
n\le c_0\sqrt{L}=c_0\rho(L,l,l').
\end{equation}
Define $\psi_j \quad (j = 0, \dots , L-1)$ 
by
\[
\psi_j (i) = 
  \left\{
   \begin{array}{lll}
N^{1/p} & {\rm if} \quad i=j\\
0 & {\rm otherwise.}
   \end{array}
   \right.
\]
Note that $\psi_j\in \mathcal{B}(L_p^N)$ and
\begin{equation}
\label{AA6}
\|J^N_{p,\infty}\psi_j\|_{L_\infty^N} = \|\psi_j\|_{L_\infty^N}=N^{1/p}.
\end{equation}
For each $u = (u_0, \dots , u_{L-1}) \in \{0,1\}^L$ define
\begin{equation}
\label{E6}
f_u = \sum_{j=0}^{L-1} u_j \psi_j.
\end{equation}
Since the functions $\psi_j$ have disjoint supports, the system
$(f_u)_{u\in\{0,1\}^L}$ satisfies condition (I). Lemma \ref{lem:9} and  
relations (\ref{D3}) and (\ref{AA6}) give
\begin{eqnarray*}
e_n^\q(J^N_{p,\infty},\mathcal{B}(L_p^N))&
\ge& \frac{1}{2}\min\big\{ \|J^N_{p,\infty} f_u-J^N_{p,\infty} f_{u'}\|_{L_\infty^N}
\,\big |\, |u|=0,\, |u'|=1\big\}\\
&=& \frac{1}{2} N^{1/p}.
\end{eqnarray*}
This proves the statement in the first case. Let
\begin{equation}
\label{AI1}
c_1=c_0/\sqrt{12}.
\end{equation}
Now we assume
\begin{equation}
\label{AI2}
 c_0 \sqrt{N}< n \le c_1 N .
\end{equation}
We set
\begin{equation}
\label{D7}
L=N, \quad  l=\lceil 2c_0^{-2}n^2N^{-1}\rceil, \quad l'=l+1.
\end{equation}
It follows from (\ref{AI2})  that $l> 2$. Moreover,
from (\ref{D7}), 
\begin{equation}
\label{D8}
n\le c_0\sqrt{l N/2}
\end{equation}
and, taking into account that $l> 2$,
$$ 
l/2 <l-1<  2c_0^{-2}n^2N^{-1},
$$
hence, by (\ref{AI1}) and (\ref{AI2}),
\begin{equation}
\label{D10}
l+1\le 3l/2< 6c_0^{-2}n^2N^{-1}\le 6c_0^{-2}c_1^2N= N/2.
\end{equation}
We have, by  (\ref{D8}), (\ref{D10}) and (\ref{D7}),
\begin{equation}
\label{D11}
n\le c_0\sqrt{l N/2} \le c_0 \min_{j=l,l+1}\sqrt{j(N-j)}\le c_0\rho(L,l,l').
\end{equation}
Now we define 
$\psi_j \in L_p^N \quad (j = 0, \dots , L-1)$ 
as
\[
\psi_j (i) = 
  \left\{
   \begin{array}{lll}
(l+1)^{-1/p}N^{1/p} & {\rm if} \quad i=j\\
0 & {\rm otherwise.}
   \end{array}
   \right.
\]
Then
\begin{equation}
\label{AA6a}
\|J^N_{p,\infty}\psi_j\|_{L_\infty^N} = \|\psi_j\|_{L_\infty^N}=(l+1)^{-1/p}N^{1/p}.
\end{equation}
Defining  the system
$(f_u)_{u\in\{0,1\}^L}$ as above, it satisfies condition (I), and
 $f_u\in \mathcal{B}(L_p^N)$ whenever $|u|=l,l+1$.
Lemma \ref{lem:9}, relations (\ref{D11}), (\ref{AA6a}), and the left and middle part of (\ref{D10}) 
give
\begin{eqnarray*}
e_n^\q(J^N_{p,\infty},\mathcal{B}(L_p^N))&
\ge& \frac{1}{2}\min\big\{ \|J^N_{p,\infty} f_u-J^N_{p,\infty} f_{u'}\|_{L_\infty^N}
\,\big |\, |u|=l,\, |u'|=l+1\big\}\\
&=& \frac{1}{2} (l+1)^{-1/p}N^{1/p} 
\ge\frac{1}{2}(6c_0^{-2}n^2N^{-1})^{-1/p}N^{1/p}\\ 
&=& \frac{c_0^{2/p}}{2\cdot 6^{1/p}}n^{-2/p}N^{2/p}.
\end{eqnarray*}
\end{proof}
The previous results together with Corollary \ref{cor:6} give
lower bounds also for arbitrary $q$. 
\begin{proposition}
\label{pro:5} Let $1\le p,q \le\infty$. There are constants $c_0,c_1>0$
such that for all $n,N\in \N$, with $n\le c_0N$ the following hold: 
If $p\le q$, then
$$
e_n^\q(J_{pq}^N,\mathcal{B}(L_p^N))\ge c_1\min\left(\left(\frac{N}{n}\right)^{2/p-2/q}
\left(\log \left(n/\sqrt{N}+2\right)\right)^{-2/q},N^{1/p-1/q}\right),
$$
 and if $p > q$, then 
$$
e_n^\q(J_{pq}^N,\mathcal{B}(L_p^N))\ge c_1(\log N+1)^{-2/q}.
$$
\end{proposition}
\begin{proof}
Fix any $\nu_1,\nu_2\in \N$ with
$$
e^{-\nu_1/8}+e^{-\nu_2/8}-e^{-(\nu_1+\nu_2)/8}\le 1/4.
$$
By Corollary \ref{cor:6},
$$
e_{(\nu_1+2\nu_2)n}^\q(J_{p,\infty}^N,\mathcal{B}(L_p^N))
\le 4\,e_{ n}^\q(J_{pq}^N,\mathcal{B}(L_p^N))
\,e_{n}^\q(J_{q,\infty}^N,\mathcal{B}(L_q^N)),
$$
therefore,
\begin{equation}
\label{AB1}
e_n^\q(J_{pq}^N,\mathcal{B}(L_p^N))
\ge 4^{-1}e_{(\nu_1+2\nu_2)n}^\q(J_{p,\infty}^N,\mathcal{B}(L_p^N)) \,
e_n^\q(J_{q,\infty}^N,\mathcal{B}(L_q^N))^{-1}.
\end{equation}
By Proposition \ref{pro:2},
\begin{equation}
\label{W1}
e_n^\q(J_{q,\infty}^{N},\mathcal{B}(L_q^N))\le 
c\,\min\left(\left(\frac{N}{n}\log \left(n/\sqrt{N}+2\right)\right)^{2/q},N^{1/q}
\right),
\end{equation}
while by Proposition \ref{pro:3}
for $n$ such that $(\nu_1+2\nu_2)n\le c_1N$ ($c_1$ the constant from Proposition 
\ref{pro:3})
\begin{eqnarray}
\label{W2}
e_{(\nu_1+2\nu_2)n}^\q(J_{p,\infty}^{N},\mathcal{B}(L_p^N))&\ge &
c_2\,\min\left(\left(\frac{N}{(\nu_1+2\nu_2)n}\right)^{2/p},N^{1/p}\right)\nonumber\\
&\ge& 
c\,\min\left(\left(\frac{N}{n}\right)^{2/p},N^{1/p}
\right).
\end{eqnarray}
We first consider the case 
$n\le \min\left(\sqrt{N},c_1 N/(\nu_1+2\nu_2)\right)$. Then 
(\ref{AB1}), (\ref{W1}), and (\ref{W2}) give  
\begin{equation}
\label{W3}
e_n^\q(J_{pq}^N,\mathcal{B}(L_p^N))\ge c N^{1/p-1/q}.
\end{equation}
In the case $\sqrt{N}<n\le c_1 N/(\nu_1+2\nu_2)$ we obtain similarly,
\begin{equation}
\label{W4}
e_n^\q(J_{pq}^N,\mathcal{B}(L_p^N))\ge 
c \left(\frac{N}{n}\right)^{2/p}
\left(\frac{N}{n}\log \left(n/\sqrt{N}+2\right)\right)^{-2/q}.
\end{equation}
For $p\le q$ relations (\ref{W3}) and (\ref{W4}) give the lower bound. In the case 
$p> q$ we note that it suffices to prove the lower bound 
 for $n=\lfloor c_1 N/(\nu_1+2\nu_2)\rfloor$. But in this case (\ref{W4}) gives 
$$
e_n^\q(J_{pq}^N,\mathcal{B}(L_p^N))\ge 
c (\log N+1)^{-2/q}.
$$
\end{proof}

To present the results in a form which emphasizes the main, polynomial parts
of the estimates and suppresses the logarithmic factors, we introduce the
following notation: For functions $a,b:\N^2\to [0,\infty)$ we write 
$a(n,N)\asymp_{\log} b(n,N)$ if there are constants 
$c_0,c_1,c_2>0$, $n_0\in \N$, $\alpha_1,\alpha_2\in \R$ such that 
$$
c_1(\log(N+n))^{\alpha_1}b(n,N)\le a(n,N)\le c_2(\log(N+n))^{\alpha_2}b(n,N)
$$
for all 
$n,N\in \N$ with $n_0\le n\le c_0 N$. In this notation, we summarize 
the results of Propositions \ref{pro:2} and \ref{pro:5} in 
\begin{theorem}
\label{theo:1} Let $1\le p,q \le\infty$. 
If $p< q$ then
$$
 e_n^\q(J_{pq}^N,\mathcal{B}(L_p^N))\asymp_{\log} 
 \min\left(\left(\frac{N}{n}\right)^{2/p-2/q},N^{1/p-1/q}\right),
$$
and if $p\ge q$, then 
$$
e_n^\q(J_{pq}^N,\mathcal{B}(L_p^N))\asymp_{\log} 1.
$$
\end{theorem}
Although the polynomial part of the order has been determined, there remains 
some room for improvements of the logarithmic factors. In the sequel we 
present some improvements of the lower bounds for particular situations. 
They involve a different way of applying 
Corollary  \ref{cor:6}, which is interesting in itself: we combine it with known 
results about summation. Furthermore, we
will use these bounds for the Sobolev case \cite{Hei03b}. We need to recall some previous
results about summation.
Define $S_N:L_p^N\to\R$ by 
$$
S_N f=\frac{1}{N}\sum_{i=0}^{N-1}f(i). 
$$
Let us recall what is known about the minimal query errors of the  $S_N$:
\begin{proposition}
\label{pro:6} Let $1\le p\le\infty$.
There are constants $c_0,c_1,c_2>0$ 
such that for all $n,N\in\N$ with $2<n\le c_0 N$,
\begin{equation}
\label{K1} c_1 n^{-1}\le e_n^\q(S_N,\mathcal{B}(L_p^N))\le c_2 n^{-1}
\quad\mbox{if}\quad 2<p\le\infty,
\end{equation}
\begin{equation}
\label{K2}
c_1 n^{-1}\le e_n^\q(S_N,\mathcal{B}(L_2^N))\le c_2 n^{-1}\log^{3/2}n\log\log n,
\end{equation}
and 
\begin{eqnarray}
\lefteqn{
c_1 \min(n^{-2(1-1/p)},n^{-2/p}N^{2/p-1})\le e_n^\q(S_N,\mathcal{B}(L_p^N))}\nonumber\\
&&
\le c_2\min(n^{-2(1-1/p)},n^{-2/p}N^{2/p-1})(\log(n/\sqrt{N}+2))^{2/p-1} \label{K3}
\end{eqnarray}
if $1\le p<2$.
\end{proposition}
The upper bound in the case $p=\infty$ is contained in
\cite{G2}, \cite{BHMT00}, 
the lower bound for $p=\infty$ is from \cite{NW}, while
(\ref{K1}) for $2<p<\infty$ and (\ref{K2}) were obtained in
\cite{Hei01}, and (\ref{K3}) is 
from \cite{HN01b}. For the case $p=2$ we need another 
upper estimate, which is more precise than  (\ref{K2}) for $n$ close to $N$.
It can be found in \cite{Hei01b}, Lemma 6.

\begin{lemma}\label{lem:6}
There is a constant $c>0$ such that for all $n,N\in\N$
with $n \le N$,
$$
e_n^\q(S_N,\mathcal{B}(L_2^N))\le cn^{-1}\l(n,N)^{3/2}\log\l(n,N),
$$
where 
$$
\l(n,N)=\log(N/n)+\log\log (n+1)+2.             
$$
\end{lemma}
\begin{proposition}
\label{pro:7} 
Let $1\le p,q\le  \infty$. There are constants $c_0,c_1>0$ such that 
for all $n,N\in\N$ with $N>4$ and $n\le c_0 N$ 
\begin{eqnarray}
e_n^\q(J_{p,q}^{N},\mathcal{B}(L_p^N))&\ge& c_1\quad \mbox{if}\quad 2<q\le\infty
\nonumber\\
e_n^\q(J_{p,2}^{N},\mathcal{B}(L_p^N))&\ge& c_1(\log \log N)^{-3/2}(\log\log\log N)^{-1}  
\nonumber\\
e_n^\q(J_{p,q}^{N},\mathcal{B}(L_p^N))&\ge& c_1(\log N)^{-2/q+1}
\quad \mbox{if}\quad 1\le q<2.
\nonumber
\end{eqnarray}
\end{proposition}
\begin{proof} The proof is similar to that of Proposition \ref{pro:5}.
Fix $\nu_1,\nu_2\in \N$ with
$$
e^{-\nu_1/8}+e^{-\nu_2/8}-e^{-(\nu_1+\nu_2)/8}\le 1/4.
$$
Since $J_{pq}^N$ is the identity, we have 
$S_N=S_N J_{pq}^N$, and we deduce from Corollary \ref{cor:6},
$$
e_{(\nu_1+2\nu_2)n}^\q(S_N,\mathcal{B}(L_p^N))
\le 4\,e_{ n}^\q(J_{pq}^N,\mathcal{B}(L_p^N))
\,e_{n}^\q(S_N,\mathcal{B}(L_q^N))
$$
which gives
\begin{equation*}
e_n^\q(J_{pq}^N,\mathcal{B}(L_p^N))
\ge 4^{-1}e_{(\nu_1+2\nu_2)n}^\q(S_N,\mathcal{B}(L_p^N)) \,
e_n^\q(S_N,\mathcal{B}(L_q^N))^{-1}.
\end{equation*}
It remains to apply Proposition \ref{pro:6} and Lemma \ref{lem:6} 
with $n=\lfloor c_0 (\nu_1+2\nu_2)^{-1} N\rfloor$.
\end{proof}
\section{Comments}
First we discuss the cost of the presented
(optimal with respect to the number of queries) algorithms 
in the bit model of computation. 
For this purpose we assume that $n$ and $N$ are powers of 2. 
The algorithm behind Proposition \ref{pro:2}
for approximating $J^N_{pq}$  is nontrivial only
for $n\ge \sqrt{N}$. In this case it uses $n$ quantum queries to 
finds all coordinates of $f\in \mathcal{B}(L_p^N)$
with absolute value not smaller than
$$
  M= c(N/n)^{2/p}\max(\log(n/\sqrt{N}),1)^{2/p}
$$
(with $c$ a concrete constant, see \cite{HN01b}). It follows from
$\|f\|_{L_p^N}\le 1$ that there are at most 
$$
NM^{-p}=\mathcal{O}(n^2 N^{-1}\max(\log(n/\sqrt{N}),1)^{-2})
$$
of them. The algorithm of
finding them, which is described and analyzed in \cite{HN01b}, needs
$\mathcal{O}(n\log N)$ quantum gates,         
$\mathcal{O}(\log N)$ qubits,  
$$
\mathcal{O}(n^2N^{-1}\max(\log(n/\sqrt{N}),1)^{-1})
$$
 measurements, and 
$\mathcal{O}(n^2N^{-1}\log N)$ classical bit operations. So the total
bit cost is $\mathcal{O}(n\log N)= \mathcal{O}(n\log n)$, hence, up to a logarithm, the 
same amount as the number of queries.

In the following table we summarize the results of this paper on the quantum 
approximation of the 
identical embeddings $J_{pq}^N:\mathcal{B}(L_p^N)\to L_q^N$ and compare them with
the respective known quantities in the classical deterministic and randomized settings.
We refer to \cite{Hei93} and the bibliography therein  for more information
on the classical settings. The respective entries of the table 
give the minimal errors, constants and logarithmic factors are suppressed. 
We always assume $n\le cN$.

\[%\hspace*{-.5cm}
\begin{array}{l|l|l|l}
\ J_{pq}^N:\mathcal{B}(L_p^N)\to L_q^N& \ \mbox{deterministic}\  & \, \mbox{random}\, & \, \mbox{quantum}\,\\ \hline 
&&&\\
1\le p< q\le \infty,\,n\le \sqrt{N}  &\, N^{1/p-1/q} & \, N^{1/p-1/q}  
& \, N^{1/p-1/q}\\
&&&\\
1\le p< q\le \infty,\ n> \sqrt{N}&\, N^{1/p-1/q} & \, N^{1/p-1/q}  
& \, \left(\frac{N}{n}\right)^{2/p-2/q}\\
&&&\\
1\le q\le p\le \infty&\, 1 & \, 1  
& \, 1\\
\end{array}
\]

We see that the quantum rate can improve 
the classical deterministic and randomized rates by a factor of 
order $N^{-1}$ (for $p=1$, $q=\infty$, and $n$ of the order of $N$).
It is this case which will lead to a speedup for Sobolev embeddings
 by an exponent 1,
see \cite{Hei03b}. 
We observe that there are also regions where
the speedup is smaller or there is no speedup at all.


\begin{thebibliography}{BHMT00}


\bibitem{Aha98}
D. Aharonov, 
Quantum computation -- a review, in: D. Stauffer (Ed.)
Annual Review of Computational Physics, vol. VI,
World Scientific, Singapore, 1998,  see also
http://arXiv.org/abs/quant-ph/9812037.

\bibitem{BHMT00}
G.~Brassard, P.~H{\o}yer, M.~Mosca, A.~Tapp, 
 Quantum amplitude amplification and estimation, 2000,
see http://arXiv.org/abs/quant-ph/0005055.

\bibitem{BBC:98}
R.~Beals,  H.~Buhrman, R.~Cleve, M.~Mosca, R.~de Wolf,
 Quantum lower bounds by polynomials, 
 Proceedings of 39th IEEE FOCS, 1998, 352-361, see also
 http://arXiv.org/abs/quant-ph/9802049.


\bibitem{EHI00}
A.~Ekert,  P.~Hayden, H.~Inamori, Basic concepts in quantum computation, 2000,
see http://arXiv.org/abs/quant-ph/0011013.

\bibitem{G2} 
L. Grover,
A framework for fast quantum mechanical algorithms,
Proceedings of the 30th Annual ACM Symposium on the Theory of Computing, 
ACM Press, New York, 1998, 53--62,  
see also http://arXiv.org/abs/quant-ph/9711043. 


\bibitem{Gru99}
J. Gruska,
Quantum Computing,
McGraw-Hill, London, 1999.


\bibitem{Hei93}
S. Heinrich,
Random approximation in numerical analysis,
in: K.~D. Bierstedt, A.~Pietsch, W.~M. Ruess, D.~Vogt (Eds.),
 Functional Analysis, Marcel {D}ekker, New York, 1993, 123 -- 171.

\bibitem{Hei01}
S. Heinrich, 
Quantum summation with an application to integration, 
Journal of Complexity 18 (2002), 1--50, see also 
http://arXiv.org/abs/quant-ph/0105116.

\bibitem{Hei01a} 
S. Heinrich, From Monte Carlo to quantum computation,
Proceedings of the 3rd IMACS Seminar on Monte Carlo
Methods MCM2001, Salzburg,
Special Issue of Mathematics and Computers in Simulation
(Guest Eds.: K. Entacher, W. Ch. Schmid, A. Uhl) 
62 (2003), 219--230.

\bibitem{Hei01b} 
S.\ Heinrich, 
Quantum integration in 
Sobolev classes, J. Complexity 19 (2003), 19--42, see also 
http://arXiv.org/abs/quant-ph/0112153.

\bibitem{Hei03b} 
S.\ Heinrich, 
Quantum Approximation II. Sobolev Embeddings, 2003

\bibitem{HKW03}
S.\ Heinrich, M.\ Kwas, H.\ Wo\'zniakowski, 
Quantum Boolean Summation with Repetitions in the
Worst-Average Setting, 2003, submitted to the Proceedings of the   
5th International Conference on Monte Carlo and Quasi-Monte Carlo Methods,
Singapore, 2002.

\bibitem{HN01a} 
S.\ Heinrich, E.\ Novak, Optimal summation and integration by deterministic, 
randomized, and quantum algorithms, in:  K.-T. Fang, F.~J. Hickernell,
H.~Niederreiter (Eds.), Monte Carlo and Quasi-Monte Carlo Methods 2000, 
Springer-Verlag, Berlin, 2002, pp. 50--62,
see also http://arXiv.org/abs/quant-ph/0105114.

\bibitem{HN01b} 
S.\ Heinrich, E.\ Novak, On a problem in quantum summation,
J. Complexity  19 (2003), 1--18, see also http://arXiv.org/abs/quant-ph/0109038.


\bibitem{KW03}
M.\ Kwas, H.\ Wo\'zniakowski,
Sharp error bounds on quantum Boolean summation in various settings, 2003,
http://arXiv.org/abs/quant-ph/0303049.

\bibitem{NW} 
A. Nayak, F. Wu,
The quantum query complexity of approximating the median and related statistics, 
STOC, May 1999, 384--393, 
see also http://arXiv.org/abs/quant-ph/9804066.

\bibitem{NC00}
M. A. Nielsen, I. L. Chuang,
Quantum Computation and Quantum Information, Cambridge
University Press, Cambridge, 2000.

\bibitem{Nov88} 
E. Novak,
Deterministic and Stochastic Error Bounds in Numerical Analysis, 
Lecture Notes in Mathematics {\bf 1349}, Springer-Verlag, Berlin, 1988. 

\bibitem{Nov01} 
E. Novak,
Quantum complexity of integration,
J. Complexity 17 (2001), 2--16,
see also http://arXiv.org/abs/quant-ph/0008124. 

\bibitem{NSW02}
E. Novak, I. H. Sloan, H. Wo\'zniakowski,
Tractability of approximation for weighted Korobov spaces on 
classical and quantum computers, 2002, see  http://arXiv.org/abs/quant-ph/0206023.

\bibitem{Pie87} A. Pietsch,
Eigenvalues and s-Numbers, 
Cambridge University Press, 1987. 

\bibitem{Pit99} 
 A. O. Pittenger,
Introduction to Quantum Computing Algorithms,
Birk\-h\"auser, Boston, 1999.

\bibitem{Pol}
D. Pollard,
Convergence of Stochastic Processes,
Springer-Verlag, New York, 1984.


\bibitem{Sho00}
P. W. Shor,
Introduction to quantum algorithms, 2000, 
see http://arXiv.org/abs/quant-ph/0005003. 

\bibitem{TW01} 
J. F. Traub, H. Wo\'zniakowski,
Path integration on a quantum computer, 2001, see
http://arXiv.org/abs/quant-ph/0109113.


\bibitem{TWW88} 
 J. F. Traub,  G. W. Wasilkowski, H. Wo\'zniakowski,
Information-Based Complexity, Academic Press, New York, 1988. 


\end{thebibliography}
\end{document}